\documentclass{webofc}
\UseRawInputEncoding
\inputencoding{latin1}
\include{Umlaute}       
\inputencoding{utf8} 
\usepackage[varg]{txfonts}   
\usepackage[utf8]{inputenc}
\usepackage{color}
\pdfoutput=1
\usepackage{float}
\newcommand*\aap{A\&A}

\newcommand*\apj{ApJ}
\newcommand*\apjl{ApJ}

\newcommand*\araa{ARA\&A}

\begin{document}
\newcommand{\hamza}[1]{\textcolor{blue}{\textbf{#1}}}
\title{Probing the role of magnetic fields in star-forming filaments: NIKA2-Pol
commissioning results toward OMC-1}
\author{\firstname{H.}~\lastname{Ajeddig}\inst{\ref{CEA}}\fnsep\thanks{\email{hamza.ajeddig@cea.fr}}
  \and \firstname{R.}~\lastname{Adam} \inst{\ref{LLR}}
  \and  \firstname{P.}~\lastname{Ade} \inst{\ref{Cardiff}}
  \and  \firstname{P.}~\lastname{Andr\'e} \inst{\ref{CEA}}
  \and \firstname{E.}~\lastname{Artis} \inst{\ref{LPSC}}
  \and  \firstname{H.}~\lastname{Aussel} \inst{\ref{CEA}}
  \and  \firstname{A.}~\lastname{Beelen} \inst{\ref{IAS}}
  \and  \firstname{A.}~\lastname{Beno\^it} \inst{\ref{Neel}}
  \and  \firstname{S.}~\lastname{Berta} \inst{\ref{IRAMF}}
  \and  \firstname{L.}~\lastname{Bing} \inst{\ref{LAM}}
  \and  \firstname{O.}~\lastname{Bourrion} \inst{\ref{LPSC}}
  \and  \firstname{M.}~\lastname{Calvo} \inst{\ref{Neel}}
  \and  \firstname{A.}~\lastname{Catalano} \inst{\ref{LPSC}}
  \and  \firstname{M.}~\lastname{De~Petris} \inst{\ref{Roma}}
  \and  \firstname{F.-X.}~\lastname{D\'esert} \inst{\ref{IPAG}}
  \and  \firstname{S.}~\lastname{Doyle} \inst{\ref{Cardiff}}
  \and  \firstname{E.~F.~C.}~\lastname{Driessen} \inst{\ref{IRAMF}}
  \and  \firstname{A.}~\lastname{Gomez} \inst{\ref{CAB}}
  \and  \firstname{J.}~\lastname{Goupy} \inst{\ref{Neel}}
  \and  \firstname{F.}~\lastname{K\'eruzor\'e} \inst{\ref{LPSC}}
  \and  \firstname{C.}~\lastname{Kramer} \inst{\ref{IRAME}}
  \and  \firstname{B.}~\lastname{Ladjelate} \inst{\ref{IRAME}}
  \and  \firstname{G.}~\lastname{Lagache} \inst{\ref{LAM}}
  \and  \firstname{S.}~\lastname{Leclercq} \inst{\ref{IRAMF}}
  \and  \firstname{J.-F.}~\lastname{Lestrade} \inst{\ref{LERMA}}
  \and  \firstname{J.-F.}~\lastname{Mac\'ias-P\'erez} \inst{\ref{LPSC}}
  \and  \firstname{A.}~\lastname{Maury} \inst{\ref{CEA}}
  \and  \firstname{P.}~\lastname{Mauskopf} \inst{\ref{Cardiff},\ref{Arizona}}
  \and \firstname{F.}~\lastname{Mayet} \inst{\ref{LPSC}}
  \and  \firstname{A.}~\lastname{Monfardini} \inst{\ref{Neel}}
  \and  \firstname{M.}~\lastname{Mu\~noz-Echeverr\'ia} \inst{\ref{LPSC}}
  \and  \firstname{L.}~\lastname{Perotto} \inst{\ref{LPSC}}
  \and  \firstname{G.}~\lastname{Pisano} \inst{\ref{Cardiff}}
  \and  \firstname{N.}~\lastname{Ponthieu} \inst{\ref{IPAG}}
  \and  \firstname{V.}~\lastname{Rev\'eret} \inst{\ref{CEA}}
  \and  \firstname{A.~J.}~\lastname{Rigby} \inst{\ref{Cardiff}}
  \and  \firstname{A.}~\lastname{Ritacco} \inst{\ref{IAS}, \ref{ENS}}
  \and  \firstname{C.}~\lastname{Romero} \inst{\ref{Pennsylvanie}}
  \and  \firstname{H.}~\lastname{Roussel} \inst{\ref{IAP}}
  \and  \firstname{F.}~\lastname{Ruppin} \inst{\ref{MIT}}
  \and  \firstname{K.}~\lastname{Schuster} \inst{\ref{IRAMF}}
  \and  \firstname{S.}~\lastname{Shu} \inst{\ref{Caltech}}
  \and  \firstname{A.}~\lastname{Sievers} \inst{\ref{IRAME}}
  \and  \firstname{C.}~\lastname{Tucker} \inst{\ref{Cardiff}}
  \and  \firstname{R.}~\lastname{Zylka} \inst{\ref{IRAMF}}
  \and  \firstname{Y.}~\lastname{Shimajiri} \inst{\ref{NAOJ},\ref{kagoshima}}
}
\institute{
  AIM, CEA, CNRS, Universit\'e Paris-Saclay, Universit\'e Paris Diderot, Sorbonne Paris Cit\'e, 91191 Gif-sur-Yvette, France
  \label{CEA}
  \and
  LLR (Laboratoire Leprince-Ringuet), CNRS, École Polytechnique, Institut Polytechnique de Paris, Palaiseau, France
  \label{LLR}
  \and
  School of Physics and Astronomy, Cardiff University, Queen’s Buildings, The Parade, Cardiff, CF24 3AA, UK 
  \label{Cardiff}
  \and
  Univ. Grenoble Alpes, CNRS, Grenoble INP, LPSC-IN2P3, 53, avenue des Martyrs, 38000 Grenoble, France
  \label{LPSC}
  \and
  Institut d'Astrophysique Spatiale (IAS), CNRS, Universit\'e Paris Sud, Orsay, France
  \label{IAS}
  \and
  Institut N\'eel, CNRS, Universit\'e Grenoble Alpes, France
  \label{Neel}
  \and
  Institut de RadioAstronomie Millim\'etrique (IRAM), Grenoble, France
  \label{IRAMF}
  \and
  Aix Marseille Univ, CNRS, CNES, LAM (Laboratoire d'Astrophysique de Marseille), Marseille, France
  \label{LAM}
  \and 
  Dipartimento di Fisica, Sapienza Universit\`a di Roma, Piazzale Aldo Moro 5, I-00185 Roma, Italy
  \label{Roma}
  \and
  Univ. Grenoble Alpes, CNRS, IPAG, 38000 Grenoble, France 
  \label{IPAG}
  \and
  Centro de Astrobiolog\'ia (CSIC-INTA), Torrej\'on de Ardoz, 28850 Madrid, Spain
  \label{CAB}
  \and  
  Instituto de Radioastronom\'ia Milim\'etrica (IRAM), Granada, Spain
  \label{IRAME}
  \and 
  LERMA, Observatoire de Paris, PSL Research University, CNRS, Sorbonne Universit\'e, UPMC, 75014 Paris, France  
  \label{LERMA}
  \and
  School of Earth and Space Exploration and Department of Physics, Arizona State University, Tempe, AZ 85287, USA
  \label{Arizona}
  \and 
  Laboratoire de Physique de l’\'Ecole Normale Sup\'erieure, ENS, PSL Research University, CNRS, Sorbonne Universit\'e, Universit\'e de Paris, 75005 Paris, France 
  \label{ENS}
  \and
  Department of Physics and Astronomy, University of Pennsylvania, 209 South 33rd Street, Philadelphia, PA, 19104, USA
  \label{Pennsylvanie}
  \and 
  Institut d'Astrophysique de Paris, CNRS (UMR7095), 98 bis boulevard Arago, 75014 Paris, France
  \label{IAP}
  \and 
  Kavli Institute for Astrophysics and Space Research, Massachusetts Institute of Technology, Cambridge, MA 02139, USA
  \label{MIT}
  \and
  Caltech, Pasadena, CA 91125, USA
  \label{Caltech}
  \and 
  National Astronomical Observatory of Japan, National Institutes of Natural Sciences, Osawa, Mitaka, Tokyo 181-8588, Japan
  \label{NAOJ}
  \and 
  Department of Physics and Astronomy, Graduate School of Science and Engineering, Kagoshima University, 1-21-35 Korimoto, Kagoshima, Kagoshima 890-0065, Japan
  \label{kagoshima}
}

\abstract{Dust polarization observations are a powerful, practical tool to probe the geometry (and to some extent, the strength) of magnetic fields in star-forming regions. In particular, Planck polarization data have revealed the importance of magnetic fields on large scales in molecular clouds. However, due to insufficient resolution, Planck observations are unable
to constrain the B-field geometry on prestellar and protostellar scales. The high angular resolution of 11.7 arcsec provided by NIKA2-Pol 1.15 mm polarimetric imaging, corresponding to $\sim$ 0.02 pc at the distance of the Orion molecular cloud (OMC), makes it possible to advance our understanding of the B-field morphology in star-forming filaments and dense cores (IRAM 30m large program B-FUN). The commissioning of the NIKA2-Pol instrument has led to several challenging issues, in particular, the instrumental polarization or intensity-to-polarization ``leakage'' effect. In the present paper, we illustrate how this effect can be corrected for, leading to reliable exploitable data in a structured,  
extended source such as
OMC-1. We present a statistical comparison between NIKA2-Pol and SCUBA2-Pol2 results in the OMC-1 region. We also present tentative evidence of local pinching of the B-field lines near Orion-KL, in the form of a new small-scale  hourglass pattern, in addition to the larger-scale hourglass already seen by other instruments such as Pol2.}
\maketitle
\section{Introduction}
Observations of nearby molecular clouds by the {\it Herschel Gould Belt Survey } have shown the crucial role of filaments in the star formation process \citep{andre2010,andre2014}, 
but the detailed fragmentation manner 
of star-forming filaments remains under debate. In parallel to {\it Herschel}, dust polarization observations from the {\it Planck satellite} have highlighted the role of magnetic fields in interstellar clouds \citep{planck2013,planck2016}. The magnetic field inferred from {\it Planck} polarization observations appears to have a regular,  structured geometry on large scales ($>>$ 0.1 pc), suggesting it is dynamically important. 
Stars form inside molecular filaments characterized by a common inner width of $\sim$0.1 pc \citep{Arzouminian2011} and the angular resolution of {\it Planck} is insufficient to probe the geometry of the B-field in the $<$0.1 pc interior of such filaments.
The IRAM 30m large program B-FUN will perform high-resolution polarization observations of a broad sample of nearby star-forming filaments imaged by {\it Herschel } \citep{andre2010}, in an effort to improve our understanding of the role of magnetic fields in core/star formation along filaments.
%
It is important to characterize the B-field at different wavelengths and high angular resolution to understand how it is involved in the star formation process \citep{crutcher2012,pattle2019,andre2019}.
%
NIKA2-Pol, the polarization channel of the NIKA2 camera \citep{NIKA2-Adam} on the IRAM 30m telescope (Pico Veleta, Spain), is under commissioning and will provide 1.15 mm continuum polarization observations with high angular resolution (11.7 arcsec). Polarimetric imaging observations are challenging due, in particular, to 
the intensity-to-polarization leakage (hereafter "leakage"), which needs to be characterized and corrected for before proper scientific observations can be made.
%
Qualitatively, the leakage affecting NIKA2-Pol data 
has the shape of a cloverleaf with a central positive lobe surrounded by a negative quadrupolar pattern \citep{AJEDDIG}.  The leakage effect can be characterized 
by observing a strong unpolarized point-like source such as 
the planet Uranus, and then subtracted from the Stokes Q, U data using a ``deconvolution'' technique described 
in detail in 
\citep{Wiesemeyer,ritac2016,aina2019}. 
%
A similar instrumental polarization effect is also present in the case of other polarimeters such as XPOL, SCUBA2-POL2, and NIKA1 \cite{Wiesemeyer, Ritacco2017, SCUBA2}.\\ 
Here, we report on the results of NIKA2-Pol commissioning observations of Orion OMC-1 that were used to characterize the leakage effect in an extended source. OMC-1 is a well documented source that has been observed by all polarimetric instruments
(see, e.g., Sect.~\ref{hourglass} below).
As part of our attempts to finalize the NIKA2-Pol commissioning, we discuss the leakage effect in OMC-1 and compare our polarization findings with recently published data (here from SCUBA2-POL2 \citep{Ward-Thompson2017}). 
As OMC-1 is one of the targets of the B-FUN large program, we also present a preliminary scientific analysis of the results (see Sect.~\ref{hourglass}), 
illustrating the great potential of NIKA2-Pol.
%
%
\label{intro}
\section{Correcting NIKA2-Pol data for instrumental polarization }
\label{leakage-corr}
\begin{figure}[H]
  \includegraphics[trim={-0.1 0.8cm 0 0},width=.9\linewidth]{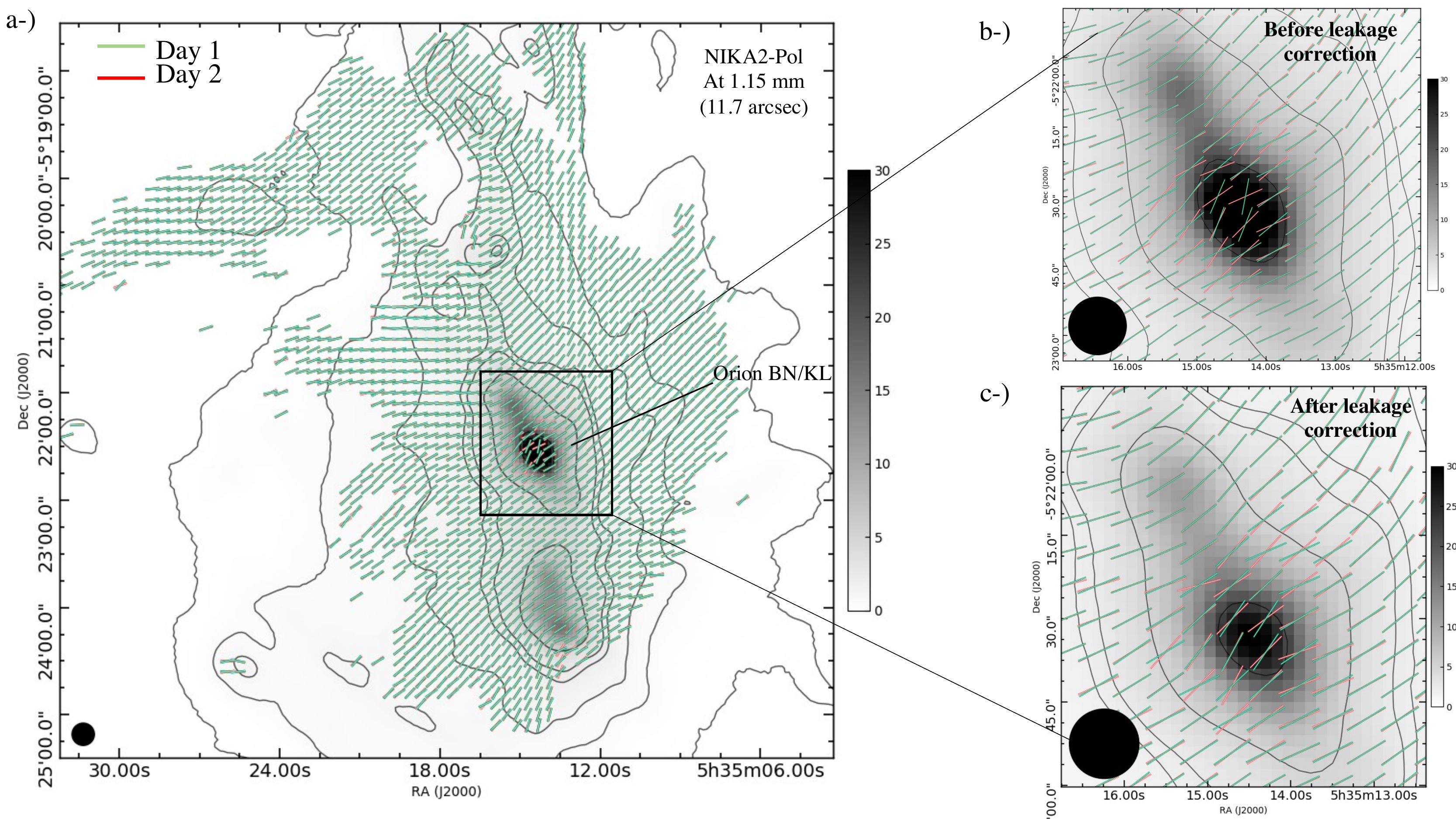}
   \caption{({\it a }): B-field vectors from NIKA2-Pol polarization data at positions of significant detections in polarized intensity (i.e., I$_{pol}$/$\sigma_{I_{pol}}>$3) on two different days, day 1 and day 2 (before leakage correction), corresponding to the green and red colors, respectively. ({\it b }): Closeup view of the vicinity of Orion-KL position overlaid with B-field vectors before correcting for instrumental polarization. ({\it c }) similar to panel {\it b} but after correcting for instrumental polarization. }%
   \label{nika2_ip}
    \end{figure}

To test the repeatability of NIKA2-Pol results, OMC-1 was observed on several nights and under good weather conditions in Nov.~2020. Figure \ref{nika2_ip} shows the difference between data taken on Nov. 12 and 15, 2020 before and after leakage correction, in panels b) and c), respectively. On the same night, the leakage pattern was observed on Uranus under similar conditions of focus and elevation, which guarantee an optimal correction of instrumental polarization (IP). Without IP correction, the OMC-1 polarization data taken on different days are in good agreement (e.g., with a root mean square difference in polarization angles of $<3^{\circ}$ ), except at the location of the compact source Orion-KL, which is strong in Stokes I but weakly polarized ($P \sim 1\%$), and thus more strongly affected by IP. Prior to IP correction, a very significant difference in polarization angles ($\sim 50^\circ$) is observed near Orion-KL between data sets taken on two different nights (see Fig.~\ref{nika2_ip}b). 
Indeed, the impact of IP is more pronounced  
in areas of the sky where the signal in Stokes I is strong 
and structured on small angular scales rather than extended and slowly varying. After subtracting the IP estimated using Uranus data observed before OMC-1, however, the polarization data from different nights agree well with each other, with a maximum difference in polarization angles reduced to $ < 20^\circ$ near Orion-KL (see Fig.~\ref{nika2_ip}c) and $\leq 2^{\circ}$ outside of Orion-KL. 
This analysis illustrates the ability of NIKA2-Pol to provide robust (repeatable), high-quality polarization maps for scientific purposes (see Fig.~\ref{omc1}a below). 
\section{Comparison of NIKA2-POL vs SCUBA2-POL2 results}
\begin{figure}[H]
   \includegraphics[trim={-0. 0.cm 0 0},width=.9\linewidth]{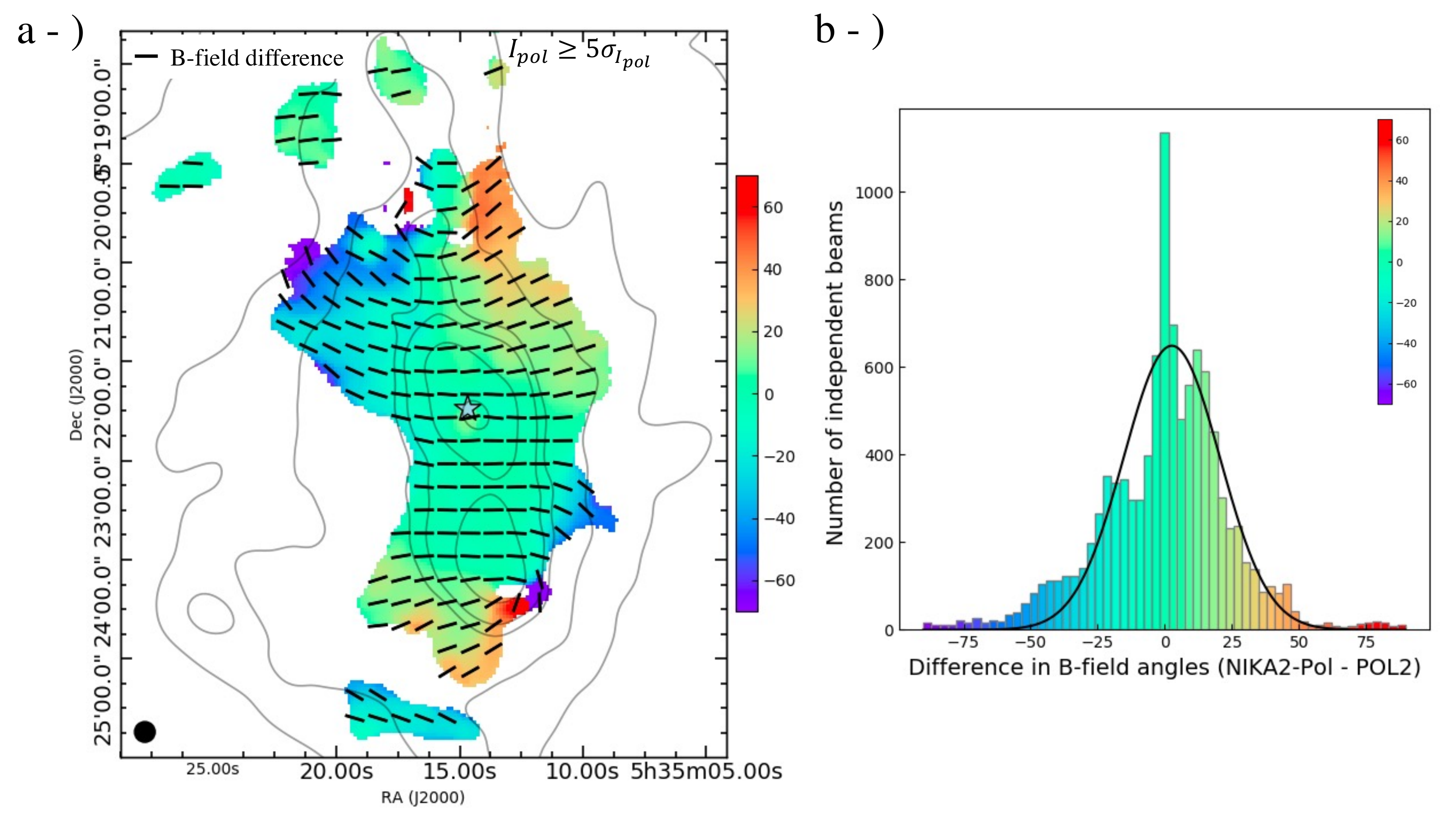}
   \caption{Differences in B-field angle between NIKA2-Pol and POL2 data in OMC-1 at the same angular resolution (14$^{\prime\prime}$) and for positions with high signal-to-noise ratio in polarized intensity 
   ($I_{pol}>5 \sigma_{I_{pol}}$ for both NIKA2-Pol and POL2). {\it a-)}: Map of the differences in B-field angle  
   (NIKA2-Pol $-$ POL2). 
   The overlaid black segments display difference vectors
   plotted every 14$^{\prime\prime}$ beam. 
   {\it b-)}: Histogram of the differences in B-field angle (NIKA2-Pol $-$ POL2). The amplitude of the difference in B-field angle between NIKA2-Pol and POL2 is less than 10$^{\circ}$ for 67\% of the positions. }
   \label{nika2_scuba}
    \end{figure}

In order to compare NIKA2-Pol polarization observations at 1.15 mm with published SCUBA2-POL2 data at 850 $\mu$m \citep{Ward-Thompson2017}, we re-projected the POL2 and NIKA2-Pol data to the same grid, and smoothed the NIKA2-Pol data to the POL2 angular resolution (14 arcsec). Figure~\ref{nika2_scuba}a shows 
map of the difference in B-field angles between POL2 and NIKA2-Pol data, where vectors represent the B-field angle difference each 14$^{\prime\prime}$.
Where the signal-to-noise ratio in polarized intensity is high  (I$_{pol}$/$\sigma_{I_{pol}}>$5) for both instruments, the NIKA2-Pol and POL2 
data are in excellent agreement, within a typical $1^{\circ}$ uncertainty 
(standard deviation of the mean difference in polarization angles -- see 
Fig.~\ref{nika2_scuba}{\it b}).
This result has been confirmed using three independent data sets from the NIKA2-Pol commissioning campaigns of December 2018, February 2020, and November 2020. 
\section{A possible new local hourglass at the Orion-KL position }
\label{hourglass}
\begin{figure}[H]
  \includegraphics[trim={-0. 0.cm 0 0},width=.9\linewidth]{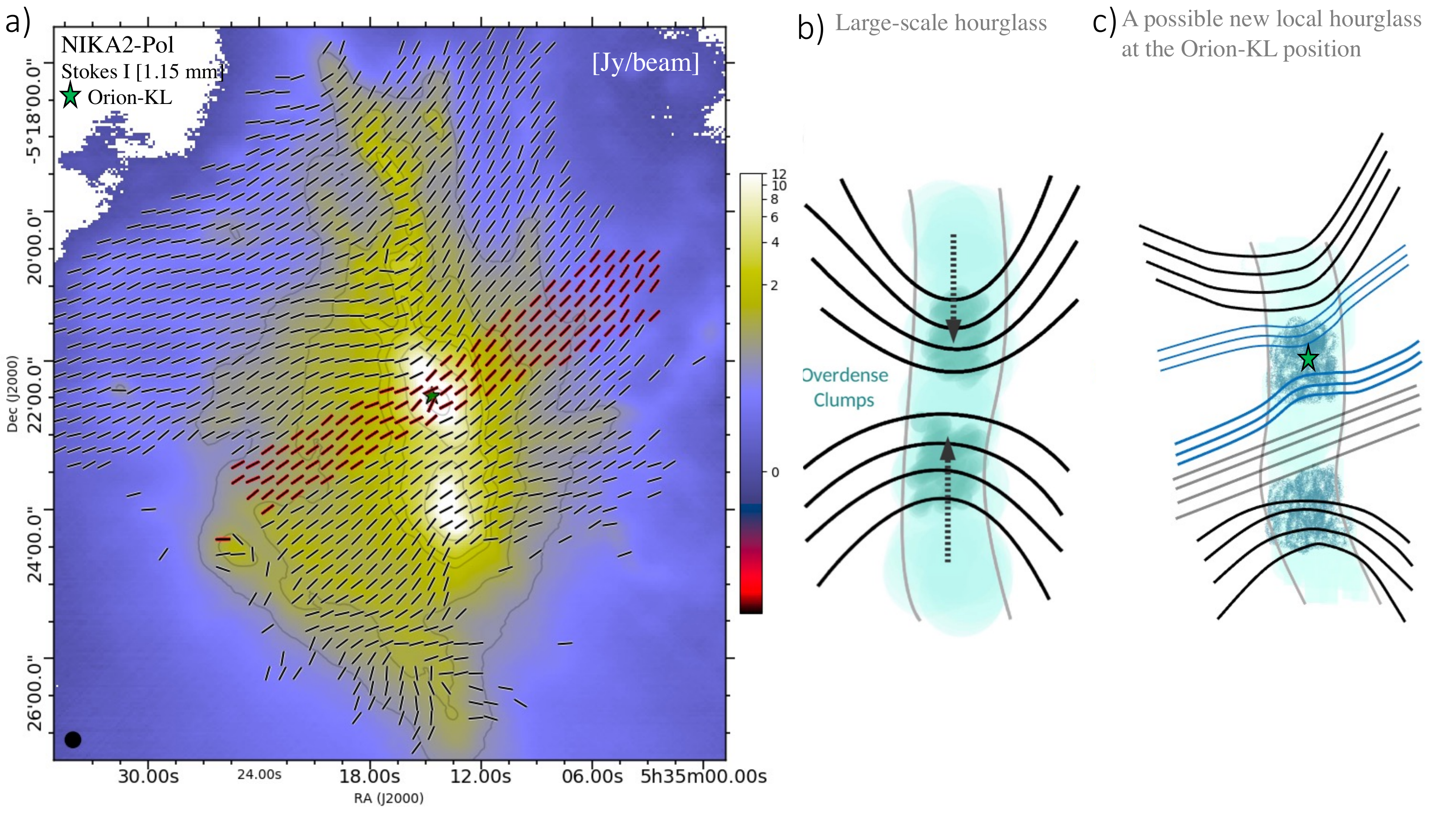}
   \caption{{\it a-)} Stokes I 1.15 mm map of OMC-1 region obtained from NIKA2-Pol polarization commissioning data, with magnetic field vectors at positions where I$_{pol}$/$\sigma_{I_{pol}}>$3. The red vectors emphasize the possible new local hourglass revealed by NIKA2-Pol. {\it b)}: Illustration of the large-scale magnetic field as seen by POL2 from Pattle et al. (2019). {\it c)} Schematic illustration of the large-scale hourglass geometry as seen with NIKA2-Pol (in black lines) and the new possible local hourglass at the position of Orion-KL revealed by the high sensitivity and resolution of NIKA2-Pol (in blue lines).    }%
   \label{omc1}
    \end{figure}

The magnetic field line geometry in OMC-1 has been under investigation for more than two decades. Schleuning (1998) \cite{s1998} reported an hourglass shape of the B-field geometry based on a 8$^\prime$ $\times$ 8$^\prime$ mapping of OMC-1 at 100 $\mu$m and 35 arcsec resolution. More recent submillimeter polarization observations with the SOFIA/HAWC+ (214 $\mu$m), SCUBA2-POL2 (850 $\mu$m) and POLKA (870 $\mu$m) polarimeters \citep{sofia,SCUBA2, polka2004} have confirmed 
this large-scale hourglass B-field geometry in OMC-1 and the weak polarization fraction around Orion-KL ($\sim$1--2\%). In particular, based on their SCUBA-POL2 study, Pattle et al. (2019) \citep{pattle2019} proposed a scenario in which the evolution of the OMC-1 region is dynamically regulated by the magnetic field. 
In this scenario, the large-scale gravitational collapse of the OMC-1 region and/or a central explosive outflow are responsible for the large-scale hourglass.
At the Orion-KL position, however, the B-field geometry was unclear. 
For instance, Vaillancourt et al. (2008) \cite{v2008} pointed out that B-field angles inferred from SHARP data at 350 $\mu$m and 450 $\mu$m differ by 25$^\circ$ and reported a drop in 
450/350\,$\mu$m polarization ratio by a factor of 2 toward Orion-KL. In addition, the B-field lines seemed to be distorted at the Orion-KL position (see Fig. 1 of \cite{v2008}; see also the ALMA observations of Cortes et al. 2021 \cite{cortes2021}). 
Thanks to the high angular resolution and sensitivity provided by NIKA2-Pol at 1.15 mm, this distortion seems to be confirmed and even looks like another hourglass, on smaller scales, centered on Orion KL, which is suggestive of local gravitational collapse at this location (see Fig.~\ref{omc1}a). 
The large-scale hourglass observed earlier is sketched in Fig.~\ref{omc1}b 
from \citep{pattle2019}, while  
the possible new local hourglass revealed by the NIKA2-Pol polarization data 
is shown and contrasted with the other hourglass in Fig.~\ref{omc1}c.
    \vspace{-1cm}
\section{Concluding remarks }
The NIKA2-Pol commissioning data have led to a better characterization of the reliability of the NIKA2 instrument in polarization mode. In this paper, we showed how  
NIKA2-Pol observations taken toward a region with extended 
and structured emission such as Orion OMC-1 
can be corrected for instrumental polarization in practice.
We achieved consistency between NIKA2-Pol data taken two years apart and also with independent observations from POL2.
The NIKA2-Pol observations confirm the presence of a large-scale hourglass pattern seen by previous polarimeters in OMC-1. In addition, the NIKA2-Pol data reveal a possible new local hourglass centered at the Orion-KL position. 
The NIKA2-Pol commissioning results obtained on OMC-1 are very promising and suggest that  high-quality scientific data will be obtained as part of the IRAM 30m large program B-FUN. 
\section*{Acknowledgements}
We would like to thank the IRAM staff for their support during the campaigns. The NIKA2 dilution cryostat has been designed and built at the Institut N\'eel. In particular, we acknowledge the crucial contribution of the Cryogenics Group, and in particular Gregory Garde, Henri Rodenas, Jean Paul Leggeri, Philippe Camus. This work has been partially funded by the Foundation Nanoscience Grenoble and the LabEx FOCUS ANR-11-LABX-0013. This work is supported by the French National Research Agency under the contracts "MKIDS", "NIKA" and ANR-15-CE31-0017 and in the framework of the "Investissements d’avenir” program (ANR-15-IDEX-02). This work has benefited from the support of the European Research Council Advanced Grant ORISTARS under the European Union's Seventh Framework Programme (Grant Agreement no. 291294). F.R. acknowledges financial supports provided by NASA through SAO Award Number SV2-82023 issued by the Chandra X-Ray Observatory Center, which is operated by the Smithsonian Astrophysical Observatory for and on behalf of NASA under contract NAS8-03060. 


\end{document}